\documentstyle[12pt,epsf]{article}
\newcommand{\bec}{\begin{center}}
\newcommand{\ec}{\end{center}}
\newcommand{\bee}{\begin{equation}}
\newcommand{\ee}{\end{equation}}

\begin{document}
\large
\begin{titlepage}
\bec
{\Large\bf  Comments on Superluminal 
Photon Tunneling \\}
\vspace*{15mm}
{\bf Yu. Malyuta \\}
\vspace*{10mm}
{\it Institute for Nuclear Research\\
National Academy of
Sciences of Ukraine\\
252022 Kiev, Ukraine\\}
{\bf e-mail: interdep@kinr.kiev.ua\\}
\vspace*{35mm}
{\bf Abstract\\}
\ec
Various expressions for transit times in 
frustrated total internal reflection are 
analysed. The incompatibility of evanescent-wave 
propagation with Einstein causality is established.

\end{titlepage}

The phenomenon of frustrated total internal
reflection illustrated in
Figure 1 has been the subject of a considerable
amount of research (see \cite{1.} and references
therein).

        The explicit expression for the transit
time in frustrated total internal reflection
has been obtained by Ghatak and Banerjee \cite{2.}.
This expression infered from the stationary
phase analysis has the form
\bee
\tau=\frac{2}{(k^{2}_{1z}+K^{2})k_{1z}K}
\Biggl(\frac{k_{1}}{v_{1}}K^{2}+\frac{k_{2}}{v_{2}}
k^{2}_{1z}\Biggr)
\ee
for $Kd \gg 1$. Here \\
$k_{1}=\frac{\omega}{c}n_{1}, \hspace*{3mm}
k_{2}=\frac{\omega}{c}n_{2}, \hspace*{3mm}
k_{1x}=k_{1}sin\theta_{i},
\hspace*{3mm} k_{1z}=k_{1}cos\theta_{i}, \hspace*{3mm}
K=\sqrt{k^{2}_{1x}-k^{2}_{2}},$ \\
$k_{1}$ and $k_{2}$ are wavenumbers in regions I and II,
$K$ is the
evanescent-wave wavenumber , $v_{1}$ and $v_{2}$
are group velocities in regions I and II,
$n_{1}$ and $n_{2}$ are refractive indexes,
$\theta_{i}$ is the incidence
angle, $\omega$ is the frequency of the incoming
wave, $d$ is the barrier width.
\begin{center}
\begin{tabular}{c}
\vspace*{-0.6cm}\\
\epsfxsize=13cm\epsffile{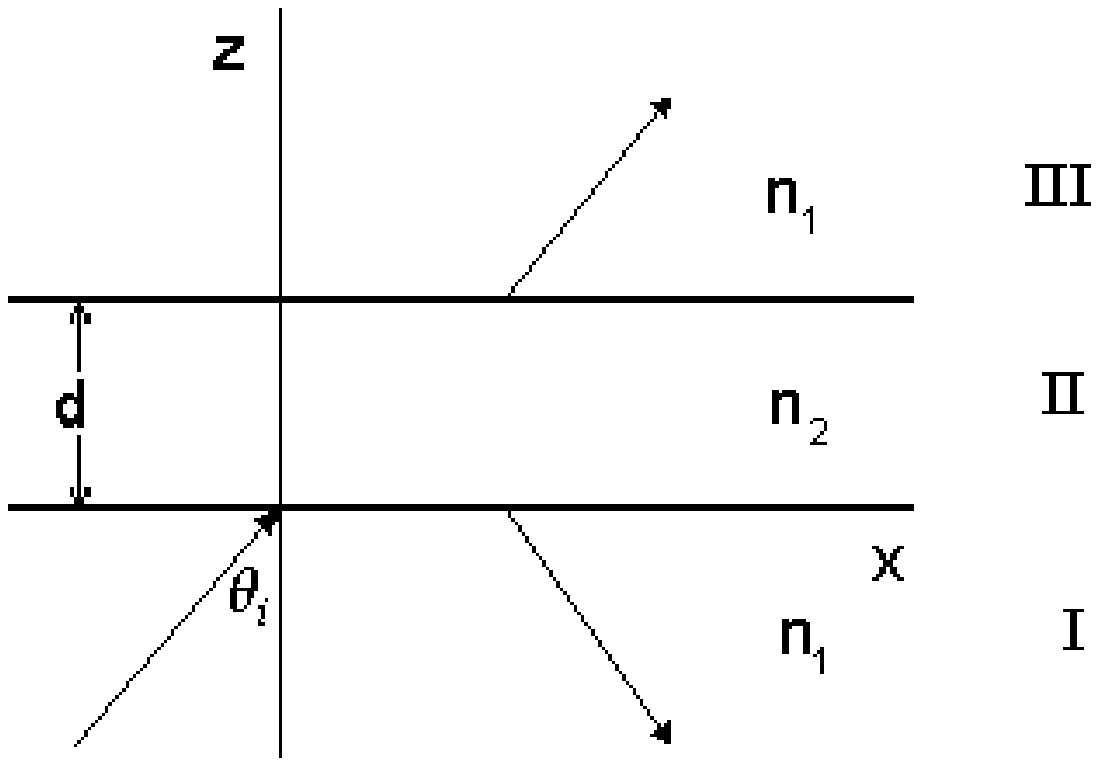}
\vspace*{0.5cm}\\
\hspace*{-2.2cm}Figure 1\\
\vspace*{0.2cm}\\
\end{tabular}
\end{center}

        Another expression for the transit time
has been recently proposed by Jakiel, Olkhovsky
and Recami \cite{3.}. This expression infered from
the analogy between photon and nonrelativistic-particle
tunneling has the form
\bee
\tau=\frac{2}{cK}
\ee
for $Kd \gg 1$.

	Note that formula (1) is valid for  all
available values of parameters ($n_{1}, \theta_{i}$).
The formula (2)
is valid only in the vicinity of the
singular point ($n_{1}\hspace*{-1mm}=
\hspace*{-2mm}\sqrt{2},
\hspace*{1mm} \theta_{i}\hspace*{-1mm}=
\frac{\pi}{4}$) because
\[\frac{2}{cK}=\frac{1}{K}
Res_{(n_{1}=\sqrt{2}, \hspace*{1mm}
\theta_{i}=\frac{\pi}{4})}
\Biggl[\frac{2}{(k^{2}_{1z}+K^{2})k_{1z}K}
\Biggl(\frac{k_{1}}{v_{1}}K^{2}+\frac
{k_{2}}{c}k^{2}_{1z}\Biggr)\Biggr] \ .\]
Therefore the expression (2) is the trivial
consequence of the formula (1).

	Superluminal photon tunneling arises the
problem of Einstein causality. To elucidate
this problem we proceed to the analysis of
evanescent-wave propagation.
According to \cite{2.}, the evanescent-wave
wavenumber $K$ satisfies the equation
\[k_{1x}^{2}-K^{2}-k_{2}^{2}=0 \ .\]
This equation is invariant under the
group $SO(1,2)$, which is the subgroup
of the group $SO(2,2)$. The group $SO(2,2)$
differs from the Lorentz group $SO(3,1)$. 
Therefore evanescent-wave propagation
is incompatible with Einstein causality.
            
\vspace*{15mm}
\hspace*{-6mm}{\bf Acknowledgement}
\vspace*{5mm}\\
The author would like to thank  V.Agranovich
for discussions.

\end{document}